\documentclass[prd,preprint,tightenlines,floatfix,showpacs,preprintnumbers,nofootinbib,eqsecnum]{revtex4}

 \usepackage[dvips,final]{graphicx}
  \usepackage{amssymb}
   \usepackage{amsmath}
    \usepackage{amsfonts}
     \usepackage{epsfig}
      \usepackage{bm}% bold math
\begin{document}

\thispagestyle{empty}
\preprint{\hbox{}} \vspace*{-10mm}

\title{Near-threshold Z-pair production\\
in the model of unstable particles with smeared mass}

\author{V.~I.~Kuksa}

\email{kuksa@list.ru}

\affiliation{Institute of Physics, Southern Federal University,
Rostov-on-Don 344090, Russia}

\author{R.~S.~Pasechnik}

\email{rpasech@theor.jinr.ru}

\affiliation{Bogoliubov Laboratory of Theoretical Physics, JINR,
Dubna 141980, Russia}

\date{\today}

\begin{abstract}
Near-threshold production of neutral boson pairs is considered
within the framework of the model of unstable particles with smeared
mass. The results of calculations are in good agreement with LEP II
data and Monte-Carlo simulations. Suggested approach significantly
simplifies calculations with respect to standard perturbative one.
\end{abstract}



\maketitle

%---------------------
\section{Introduction}
%---------------------

Theoretical description of the near-threshold $Z$- and $W$-pair
production in channels $e^+e^-\rightarrow ZZ, WW$ became actual
since LEP II operation has started \cite{1}-\cite{8}. Now the
interest to the boson-pair production remains topical as far as the
boson intermediate states enter the high energy cascade processes at
future hadron and linear colliders \cite{9}-\cite{12}. At threshold
energy $\sqrt{s}\approx 2M_{Z,W}$ the finite-width effects (FWE)
play a significant role in the boson-pair production. In the stable
particle approximation (i.e. having bosons with zero widths) the
on-shell boson-pair production has a non-physical threshold
$\sqrt{s_{th}}=2M$ which turns out to be eliminated (smeared) by
taking into account of FWE. Usually it is fulfilled with help of the
dressed boson propagator in processes like $e^+e^-\rightarrow
ZZ\rightarrow 4f$ (Double Pole Approximation).

The calculation of the total cross-section $\sigma(e^+e^-\rightarrow
4f)$ including high-order perturbative corrections is very
complicated problem and usually carried out by Monte-Carlo
simulations \cite{6,8}. The so-called semi-analytical approach (SAA)
significantly reduces the complexity of the calculations and gives
the cross-sections in a quite compact analytical form
\cite{3}-\cite{5}. This approach is based on the approximate
factorization of the cross-section. In order to simplify the
perturbative calculations and effectively incorporate the higher
order corrections the effective theory of unstable particles (UP)
was explored in \cite{11,12}.

In this paper we consider $Z$-pair production within the framework
of unstable particles model with a smeared mass \cite{13,14}. In the
model UP is described as on-shell state having variable
(smeared) mass in the vicinity of the threshold. The inclusive
process $e^+e^-\rightarrow ZZ,WW\rightarrow all(4f)$ in Double Pole
Approximation (DPA) is described as a boson-pair production process
$e^+e^-\rightarrow ZZ,WW$ where bosons are on the fuzzed (smeared)
mass-shell. This treatment is in close analogy with SAA,
but our model contains new element - the smearing (spreading) of mass, which is directly connected with instability (see Section 2). In contrast to SAA or convolution method the model leads to exact factorization in wide class of processes \cite{15,16}. As was shown in \cite{16}, the model predictions at high energies have to coincide with standard perturbative results with very high accuracy. In this paper we have got the same coincidence while the
calculations are much simpler.

The paper organized as follows. In Section 2 we briefly describe the
UPs model and derive the double convolution formula for the $Z$-pair
production cross-section. The strategy of calculations and results
including the one-loop EW corrections are discussed in Section 3. We
stressed out that our results precisely coincide with the
corresponding Monte-Carlo simulation. In Section 4 we make the
conclusion concerning the applicability of suggested method.

%---------------------------------------------
\section{Cross-section of $Z$-pair production
in the model of unstable particles with smeared mass}
%---------------------------------------------

The model of UP with smeared mass is based on fundamental relation between lifetime of unstable state and spreading of energy level. As was noted by Matthews and Salam \cite{16a}, an account of instability in wave function of UP leads to the spreading (smearing) of its mass. A wave function of UP in their rest system with an account of decay width $\Gamma =1/\tau$ can be written in terms of its Fourier transform:
\begin{equation}\label{2.0}
 \psi(t)=\exp\{iMt-\Gamma |t|/2\}\,\,\,\rightarrow\,\,\,\frac{\Gamma}{2\pi}\int\frac{\exp\{-imt\}}{(m-M)^2+\Gamma^2/4}dm.
\end{equation}
Right-hand part of (\ref{2.0}) may be interpreted as describing a distribution of mass values, with a spread, $\delta m$, related to the mean life $\delta \tau=1/\Gamma$, by an uncertainty relation $\delta m \delta \tau \sim 1$.
The generalization of the field wave function (\ref{2.0}) can be represented in the form \cite{14}:
\begin{equation}\label{2.1}
 \Psi(x)=\int\Psi(x,\mu)\omega(\mu)d\mu,
\end{equation}
where $\Psi(x,\mu)$ is standard spectral component defining a particle
with fixed mass squared $m^2=\mu$ in the stable particle
approximation (SPA) and $\omega(\mu)$ is some weight function
formed by self-energy interactions of UP with vacuum fluctuations
and decay products. This function describes the smeared (fuzzed)
mass-shell of UP. So, the smearing of mass is caused, on the one hand by instability according to formal relation (\ref{2.0}) and on the other hand by stochastic self-energy type interaction of unstable system with vacuum fluctuations.

The transition amplitude of the process with one UP is factorized as
$A(\mu)=A^{st}(\mu)\omega(\mu)$ \cite{14}, where $A^{st}(\mu)$ is
the standard amplitude in SPA. Then, the differential probability of
transition $e^+e^-\rightarrow Z(\mu_1)Z(\mu_2)$ is
\begin{equation}\label{2.2}
 dP(\mu_1,\mu_2)=|A^{st}(\mu_1,\mu_2)|^2\rho(\mu_1)\rho(\mu_2)\,d\mu_1
 d\mu_2,
\end{equation}
where $\rho(\mu)=|\omega(\mu)|^2$ is the probability density of the
mass squared distribution. As a result we get the double convolution
formula for the cross-section
\begin{equation}\label{2.3}
 \sigma(e^+e^-\rightarrow
 ZZ)=\int_{\mu_0}^{s}\int_{\mu_0}^{(\sqrt{s}-\sqrt{\mu_1})^2}
 \sigma^{st}(e^+e^-\rightarrow
 Z(\mu_1)Z(\mu_2))\rho(\mu_1)\rho(\mu_2)\,d\mu_1d\mu_2,
\end{equation}
where $\mu_0\sim m_f$ is the threshold,
$\sigma^{st}(e^+e^-\rightarrow Z(\mu_1)Z(\mu_2))$ is the
cross-section of $Z(\mu_1)$ and $Z(\mu_2)$ pair production in SPA.
As usual, we introduce the factor 1/2 to take into account the
"integral" identity of $Z(\mu_1)$ and $Z(\mu_2)$. This factor is
necessary to satisfy the stable particle limit
\begin{equation}\label{2.4}
 \rho(\mu) \rightarrow \delta(\mu -M^2),\,\,\,\sigma (e^+e^-\rightarrow
 ZZ)\rightarrow \sigma^{st}(e^+e^-\rightarrow Z(M)Z(M)).
\end{equation}
and to describe correctly the exclusive processes such as
$e^+e^-\rightarrow f_1\bar{f}_1 f_2\bar{f}_2$ and $e^+e^-\rightarrow
f_1\bar{f}_1 f_1\bar{f}_1$.

The process under consideration is given by elementary Born
cross-section in the standard form
\begin{equation}\label{2.5}
 \sigma^{st}(e^+e^-\rightarrow Z(\mu_1)Z(\mu_2))=
 \frac{g^4(1+6c^2+c^4)}{2^{10}\pi s\cos^4{\theta_W}}
 \bar{\lambda}(\mu_1,\mu_2;s)f(\mu_1,\mu_2;s),
\end{equation}
where $c=1-4\sin^2{\theta_W}$ and $g$ is the weak coupling constant.
The functions $\bar{\lambda}(\mu_1,\mu_2;s)$ and $f(\mu_1,\mu_2;s)$
are defined by the following expressions
\begin{equation}\label{2.6}
 \bar{\lambda}(\mu_1,\mu_2;s)=\left[1-2\,\frac{\mu_1+\mu_2}{s}+
 \frac{(\mu_1-\mu_2)^2}{s^2}\right]^{1/2}
\end{equation}
and
\begin{equation}\label{2.7}
 f(\mu_1,\mu_2;s)=-1+\frac{s^2+(\mu_1+\mu_2)^2}
 {s(s-\mu_1-\mu_2)\bar{\lambda}(\mu_1,\mu_2;s)}
 \arctan{\frac{s\bar{\lambda}(\mu_1,\mu_2;s)}{s-\mu_1-\mu_2}}.
\end{equation}

The probability density $\rho(\mu)$ was defined in Ref.~\cite{14} by
three different ways. We choose it in traditional Lorentzian form
\begin{equation}\label{2.8}
 \rho(\mu)=\frac{1}{\pi}\frac{\sqrt{\mu}\,\Gamma_Z(\mu)}
 {(\mu-M^2_Z)^2+\mu\,\Gamma^2_Z(\mu)},
\end{equation}
where $\Gamma_Z(\mu)$ is $\mu$-dependent total width of $Z$-boson.
The expressions (\ref{2.3}) -- (\ref{2.8}) define the cross-section
of two-boson production and, consequently, the inclusive
$4f$-production in DPA.
The results in close analogy with (\ref{2.3}) and (\ref{2.8}) arise when instability is described by decay factor $\Gamma |t|$ in operator function of final $Z$ -boson states. This effect was described in \cite{16c} as "fuzzy mass shell" for Majorana neutrinos.

To get the cross-section of exclusive
$4f$-production we represent the total width $\Gamma_Z(\mu)$ as
\begin{equation}\label{2.9}
 \Gamma_Z(\mu)=\sum_{k}\Gamma(Z(\mu)\rightarrow f_k\bar{f}_k).
\end{equation}
Then, we can represent the exclusive DPA cross-section in the form
\begin{eqnarray}\nonumber
 \sigma_{DP}(e^+e^-\to Z(\mu_1)Z(\mu_2)\to f_k\bar{f}_k\,
 f_i\bar{f}_i)&=&\int_{\mu_0}^{s}\int_{\mu_0}^{(\sqrt{s}-\sqrt{\mu_1})^2}
 \sigma^{st}(e^+e^-\rightarrow Z(\mu_1)Z(\mu_2))\times\\
 &&{}(2-\delta_{ik})\rho^k_Z(\mu_1)\rho^i_Z(\mu_2)\,d\mu_1d\mu_2,
\label{2.10}
\end{eqnarray}
where the partial distribution $\rho^k_Z(\mu)$ is defined as
\begin{equation}\label{2.11}
 \rho^k_Z(\mu)=\frac{1}{\pi}\,\frac{\sqrt{\mu}\,\Gamma(Z(\mu)\rightarrow
 f_k\bar{f}_k)}{(\mu-M^2_Z)^2+\mu\Gamma^2_Z(\mu)}.
\end{equation}
Note, the combinatorial factor $2-\delta_{ik}$ together with above
mentioned factor 1/2 takes into account the identity/nonidentity of
fermion pairs in the final state.

The expression (\ref{2.10}) has a close analogy with one obtained in
the semi-analytical approach \cite{3}, but the methodological status
of these results are different. Suggested approach is based on the fundamental relation between instability and smearing of energy level (i.e. mass of UP).
In the framework of the model, Eq.~(\ref{2.10}) is derived from the exclusive cross-section (\ref{2.3}), which directly follows from the structure of the model \cite{14}. The same form of exclusive cross-section arises in
standard approach (SAA) as an approximation, which is usually called
the "narrow-width approximation" \cite{17}. Within the
framework of UPs model \cite{14} the factorization at tree level is
exact due to specific model propagator of UP as it was shown in
Refs. \cite{15,16}. The higher order electro-weak corrections, which
play a significant role in the considering high-energy process
\cite{1} -- \cite{12}, will be discussed in the next section.

%-------------------------------------------------------------
\section{Analysis of results}
%-------------------------------------------------------------

The Born cross-section of $ZZ$-pair production for $Z$-boson states
with fixed mass (stable particle approximation) and smeared mass
(model state of UP) are represented in Fig. 1 by dashed and solid
lines, respectively. From this figure one can see that the FWE give
a significant contribution at the energy interval $2M_Z-10\,
\mathrm{GeV}\lesssim \sqrt{s}\lesssim 2M_Z+10\,\mathrm{GeV}$, where
the threshold smearing is noticeable. At higher energies these two
lines asymptotically approach each other. Both the curves
significantly exceed the LEP data because the higher order
electro-weak corrections give large contribution and have to be
taken into consideration \cite{3}.

%-------------------------------------------------------------
\begin{figure}[h!]
\centerline{\epsfig{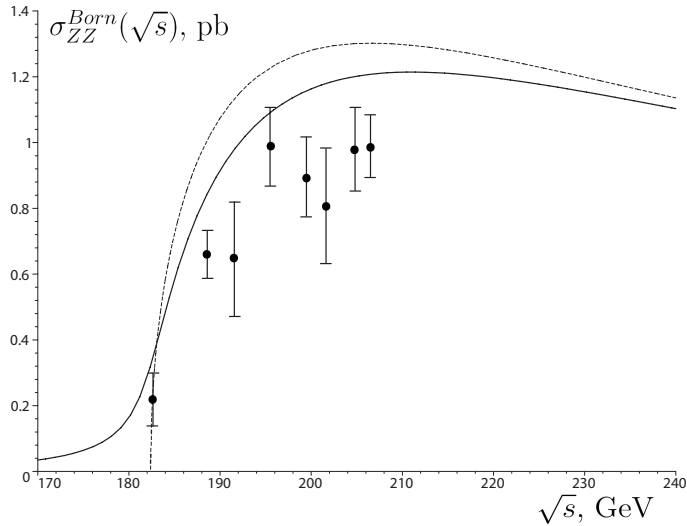}} \caption{Born $ZZ$
cross-section in stable particle approximation (dashed line) and in
the model of UP with smeared mass (solid line).} \label{fig:Born}
\end{figure}
%-------------------------------------------------------------

The strategy of calculations with taking into account of the higher
order corrections is in the following. The self-energy corrections
to the final states of unstable $Z$ are incorporated in definition
of the model wave function (\ref{2.1}), i.e. in the weight function
$\omega(\mu)$ or probability density $\rho(\mu)$. The principal part
of vertex corrections is taken into account by using the weak
running coupling $\alpha(M_Z)=1/127.9$ \cite{18}. The corrections to
the initial and intermediate states of electron and positron caused
by Initial State Radiation (ISR) and virtual radiation in the
one-loop approximation were calculated before in Ref. \cite{3}, and
we just follow the same procedure.

The model cross-section including above mentioned corrections is
represented in Fig. 2 (the solid line) together with the result of
Monte-Carlo simulation (the dashed line) and LEP data points
\cite{19}. Both results are consistent with the data within the
error bars. We note that results of Monte-Carlo simulation and our
model calculations coincide one with another with very high
precision. From this result it follows that the contribution of non-factorizable corrections in the considered energy range is small. So, we can applied our approach to the process $e^{+}e^{-}\to W^{+}W{-}$ in the same energy range. The lines start to differ slightly at energies larger
than that of the available data, i.e. at
$\sqrt{s}>200\,\mathrm{GeV}$. Note, that the difference between model and Monte-Carlo curves is an order of differences between results of various Monte-Carlo calculations.

%-------------------------------------------------------------
\begin{figure}[h!]
\centerline{\epsfig{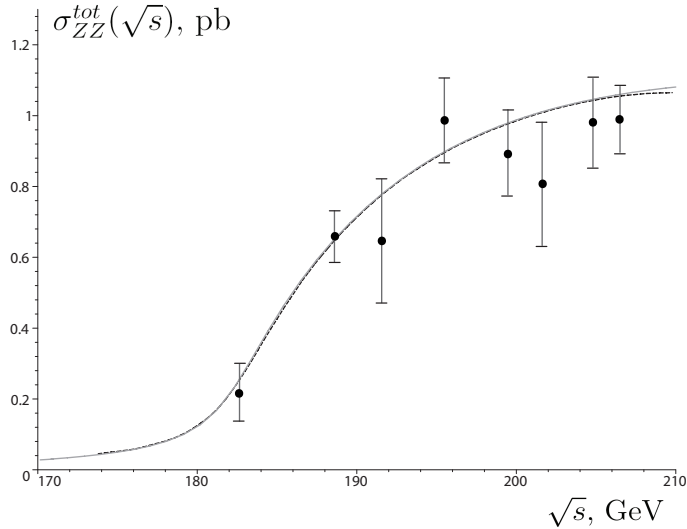}} \caption{Total $ZZ$
cross-section obtained with Monte-Carlo simulations (dashed line)
and in the model of UP with smeared mass (solid line).}
\label{fig:total}
\end{figure}
%-------------------------------------------------------------

The exclusive processes $e^+e^-\rightarrow q_i\bar{q}_iq_k\bar{q}_k$
at tree level are described by Eq.~(\ref{2.10}).
For instance, it follows from Eq.~(\ref{2.10}) that
\begin{equation}\label{2.12}
 \frac{\sigma(e^+e^-\rightarrow q_i\bar{q}_iq_k\bar{q}_k)}{\sigma(e^+e^-\rightarrow
 q_i\bar{q}_iq_i\bar{q}_i)}\approx 2,
\end{equation}
and this result is in agreement with Monte-Carlo simulations
\cite{8}. It should be noted that the model assumes to account for non-factorizable corrections caused by interaction of initial states and final decay products of $Z$ -bosons. In this case $Z$ -boson in an intermediate state is described by model propagator \cite{15,16} and the convolution structure of cross-section is destructed. As was noted, such a correction is small in considered energy range.

%-----------------------
\section{Conclusion}
%-----------------------

The calculation of $\sigma(e^+e^-\rightarrow 4f)$ together with the
higher order electro-weak corrections is very complicated and
longstanding problem intensified by nonfactorizable diagrams. Since
the cross-sections are unavailable in an analytical form, the
various approximations and Monte-Carlo simulations are usually
applied. The semi-analytical approach allows to represent the
exclusive cross-section in a compact analytical form and
significantly simplifies inclusion of higher order corrections
\cite{3}.

We have calculated the inclusive cross-section within the framework
of the model of unstable particles with a smeared (random) mass.
This approach has close analogy to the semi-analytical and
convolution methods and, as a rule, gives the same results. However,
our results are based on the simple model which explicitly describes the instability by smearing of mass and gives  the cross-section and decay width in  the convolution form. Moreover, as it was shown in Refs.~\cite{15,16}, the model provides an exact factorization in the cases when UP is in the
intermediate state. So, this model can be treated as a
methodological base for the SAA and CM. As was noted above, these approaches are valid for the discussed case giving quite accurate results.

The results of the model are in accordance with the experimental
data and coincide with Monte-Carlo results with very high accuracy.
In contrast to standard approach based on PT, the discussed method
gives simple tools for calculations. It assumes an account of high-order corrections which in most cases do not destroy the convolution structure of cross-section and decay width.

\section*{Acknowledgments}

We would like to thank V. Beylin and G. Vereshkov for fluent
discussions and correspondence. This work was supported in part by
Grants RFBR 06-02-16215 and 07-02-91557.

\end{document}